\documentclass[sigconf]{acmart}

\AtBeginDocument{%
  }

\copyrightyear{2026}
\acmYear{2026}
\setcopyright{cc}
\setcctype{by}
\acmConference[SIGIR '26]{Proceedings of the 49th International ACM SIGIR
  Conference on Research and Development in Information Retrieval}%
  {July 20--24, 2026}{Melbourne, VIC, Australia.}
\acmBooktitle{Proceedings of the 49th International ACM SIGIR Conference on
  Research and Development in Information Retrieval (SIGIR '26),
  July 20--24, 2026, Melbourne, VIC, Australia}
\acmISBN{979-8-4007-2599-9/2026/07}
\acmDOI{10.1145/3805712.3809722}
\usepackage{amsmath}
\usepackage{algorithm}
\usepackage{algorithmic}
\usepackage{booktabs}
\usepackage{multirow}
\usepackage{graphicx}
\usepackage{xcolor}
\usepackage{balance}
\usepackage{tikz}
\usepackage{pgfplots}
\usepackage{pifont}  
\pgfplotsset{compat=1.18}
\usetikzlibrary{positioning,arrows.meta,shapes.geometric,calc,patterns}

\definecolor{oiblue}{RGB}{0,114,178}
\definecolor{oiorange}{RGB}{230,159,0}
\definecolor{oigreen}{RGB}{0,158,115}
\definecolor{oired}{RGB}{213,94,0}
\definecolor{oipurple}{RGB}{204,121,167}
\definecolor{oiyellow}{RGB}{240,228,66}
\definecolor{oicyan}{RGB}{86,180,233}
\definecolor{oigray}{RGB}{128,128,128}

\newcommand{\methodname}{\textsc{ReaLM-Retrieve}}

\newcommand{\eg}{\textit{e.g.}}
\newcommand{\etal}{\textit{et al.}}

\begin{document}

\title{When to Retrieve During Reasoning: Adaptive Retrieval for Large Reasoning Models}

\author{Dongxin Guo}
\authornote{Both authors contributed equally to this research.}
\authornote{Corresponding author.}
\orcid{0009-0000-2388-1072}
\affiliation{%
  \institution{The University of Hong Kong}
  \city{Hong Kong}
  \country{China}}
\affiliation{%
	\institution{Brain Investing Limited}
	\city{Hong Kong}
	\country{China}}
\email{bettyguo@connect.hku.hk}

\author{Jikun Wu}
\authornotemark[1] 
\orcid{0000-0002-2327-4157}
\affiliation{%
  \institution{Stellaris AI Limited}
  \city{Hong Kong}
  \country{China}}
\affiliation{%
  \institution{Brain Investing Limited}
  \city{Hong Kong}
  \country{China}}
\email{hk950014@connect.hku.hk}

\author{Siu Ming Yiu}
\orcid{0000-0002-3975-8500}
\affiliation{%
  \institution{The University of Hong Kong}
  \city{Hong Kong}
  \country{China}}
  \affiliation{%
  	\institution{Brain Investing Limited}
  	\city{Hong Kong}
  	\country{China}}
\email{smyiu@cs.hku.hk}

\begin{abstract}
Large reasoning models such as DeepSeek-R1 and OpenAI o1 generate extended chains of thought spanning thousands of tokens, yet their integration with retrieval-augmented generation (RAG) remains fundamentally misaligned. Current RAG systems optimize for providing context \emph{before} reasoning begins, while reasoning models require evidence injection \emph{during} multi-step inference chains. We introduce \methodname{}, a reasoning-aware retrieval framework that addresses this mismatch through three key innovations: (1) a \emph{step-level uncertainty detector} that identifies knowledge gaps at reasoning-step granularity rather than token or sentence level; (2) a \emph{retrieval intervention policy} that learns when external evidence maximally benefits ongoing reasoning; and (3) an \emph{efficiency-optimized integration} mechanism that reduces per-retrieval overhead by 3.2$\times$ compared to naive integration. Experiments on MuSiQue, HotpotQA, and 2WikiMultiHopQA demonstrate that \methodname{} achieves on average 10.1\% absolute improvement in answer F1 over standard RAG (range: 9.0--11.8\% across the three benchmarks) while reducing retrieval calls by 47\% compared to fixed-interval approaches like IRCoT (all improvements significant at $p<0.01$, paired bootstrap). On the challenging MuSiQue benchmark requiring 2--4 hop reasoning, our method achieves 71.2\% F1 with an average of only 1.8 retrieval calls per question. Analysis shows that \methodname{} also improves retrieval quality itself, achieving 81.3\% Recall@5 with consistently higher precision and MRR than fixed-interval baselines on supporting evidence, establishing new state-of-the-art efficiency-accuracy trade-offs for reasoning-intensive retrieval tasks.
\end{abstract}



\maketitle

\section{Introduction}
\label{sec:intro}

The emergence of large reasoning models (LRMs) marks a major shift in language model capabilities. Models such as DeepSeek-R1~\cite{deepseek2025r1}, OpenAI o1~\cite{openai2024o1systemcard}, and QwQ~\cite{qwen2024qwq} generate extended chains of thought spanning 12,000--25,000 tokens, exhibiting emergent behaviors including self-verification, backtracking, and strategy switching. Unlike standard chain-of-thought (CoT) prompting~\cite{wei2022chain}, these models are trained via reinforcement learning to develop genuine multi-step reasoning capabilities that cannot be replicated through prompting alone.

However, a critical limitation persists: reasoning models excel at \emph{reasoning with known facts} but struggle when \emph{external factual knowledge} is required. Recent analyses reveal that reasoning models show \emph{degraded} factual accuracy compared to their non-reasoning counterparts on fact-seeking tasks~\cite{huang2025thinkingloud}, with increased hallucination rates on benchmarks like SimpleQA. This creates a fundamental opportunity for retrieval augmentation; yet existing RAG systems are fundamentally misaligned with reasoning model architectures.

The core mismatch is temporal: current RAG pipelines optimize for providing relevant context before generation begins~\cite{gao2024ragsurvey, lewis2020rag}. This ``retrieve-then-generate'' paradigm assumes a single retrieval step suffices. For reasoning models generating thousands of tokens across multiple logical steps, this assumption fails catastrophically. A reasoning chain may begin with sufficient context but encounter knowledge gaps mid-inference, gaps that cannot be anticipated at query time.

Iterative retrieval methods such as IRCoT~\cite{trivedi2023ircot}, FLARE~\cite{jiang2023flare}, and Self-RAG~\cite{asai2024selfrag} address this by interleaving retrieval with generation. Yet these methods were designed for standard language models, not reasoning architectures, and exhibit fundamental incompatibilities: \textbf{(1) Granularity mismatch}: Existing methods trigger retrieval at token level (FLARE) or sentence level (IRCoT), while reasoning models operate at reasoning step granularity (logical units spanning multiple sentences). \textbf{(2) Signal unavailability}: Methods relying on token probabilities (FLARE) or internal attention states (DRAGIN~\cite{su2024dragin}) cannot operate with completion-only models like o1. \textbf{(3) Efficiency collapse}: Fixed-interval retrieval (IRCoT) becomes prohibitively expensive for reasoning chains spanning thousands of tokens, adding 2--5 seconds per retrieval call.

We propose \methodname{}, a reasoning-aware retrieval framework that addresses these challenges through principled integration of retrieval into reasoning model inference. Our approach makes three key contributions:

\textbf{Contribution 1: Step-Level Uncertainty Detection.} We propose a novel uncertainty quantification method operating at reasoning-step granularity. Unlike token-level entropy or sentence-level confidence, our Reasoning Step Uncertainty Score (RSUS) identifies when a logical inference requires external knowledge versus when reasoning should continue with existing context.

\textbf{Contribution 2: Learned Retrieval Intervention Policy.} We formalize reasoning-retrieval integration as a sequential decision problem and learn an intervention policy $\pi(a_t|s_t)$ that decides whether to retrieve, what query to formulate, and how to integrate retrieved content, all conditioned on the current reasoning state.

\textbf{Contribution 3: Efficiency-Optimized Integration.} We develop implicit compression and speculative caching mechanisms that reduce per-retrieval overhead by 3.2$\times$ compared to naive integration (from 2.1s to 0.66s per call), enabling practical deployment of reasoning-retrieval systems with 1.33$\times$ lower end-to-end latency than fixed-interval approaches.

Experiments on three multi-hop QA benchmarks show that \methodname{} achieves substantial improvements in both accuracy and efficiency. On MuSiQue, we achieve 71.2\% F1 with 1.8 average retrieval calls compared to IRCoT's 65.4\% F1 with 3.4 calls (improvement significant at $p<0.01$). Our method establishes a new Pareto frontier for reasoning-retrieval trade-offs, demonstrating that fewer, better-timed retrievals outperform frequent, fixed-interval retrievals.

\section{Related Work}
\label{sec:related}

\subsection{Retrieval-Augmented Generation}
RAG systems~\cite{borgeaud2022retro, guu2020realm, lewis2020rag} augment language models with external knowledge retrieved from document collections. The standard approach involves a single retrieval step before generation. Dense retrieval methods using dual encoders~\cite{karpukhin2020dpr, xiong2021ance} or late interaction architectures~\cite{khattab2020colbert, santhanam2022colbertv2} have substantially improved retrieval quality. Recent evaluation frameworks~\cite{es2024ragas} enable principled assessment of retrieval-generation pipelines. Multi-vector retrieval engines such as PLAID~\cite{santhanam2022plaid} and WARP~\cite{scheerer2025warp} optimize late-interaction retrievers, achieving 3--41$\times$ speedups. Our work complements these advances by optimizing when to invoke retrieval during reasoning.

\subsection{Iterative and Adaptive Retrieval}
Recent work explores iterative retrieval strategies. IRCoT~\cite{trivedi2023ircot} interleaves retrieval with chain-of-thought reasoning, retrieving after each sentence. While effective for short generation (achieving 15 F1-point improvements on 2WikiMultiHopQA), fixed-interval retrieval becomes impractical for reasoning chains spanning thousands of tokens. FLARE~\cite{jiang2023flare} triggers retrieval when token probability falls below a threshold, but requires token-level probabilities unavailable from completion-only models. DRAGIN~\cite{su2024dragin} uses attention entropy but requires internal states. Self-RAG~\cite{asai2024selfrag} learns special tokens through fine-tuning, achieving strong performance but requiring full model fine-tuning impossible for proprietary models. REPLUG~\cite{shi2024replug} treats models as black boxes but operates at query-level without mid-reasoning intervention. A separate line of work makes the retrieval decision \emph{at the query level} based on question complexity: Adaptive-RAG~\cite{jeong2024adaptiverag} routes queries to no-retrieval, single-retrieval, or multi-retrieval strategies, and Open-RAG~\cite{islam2024openrag} learns reflection tokens that determine retrieval necessity per query. Most closely related to our setting, the concurrent work of Hashemi~\etal~\cite{hashemi2026costaware} (``Dynamic Search-R1'') extends Search-R1 with a cost-aware RL objective that adapts retrieval \emph{depth} (the number of documents per sub-query); their orthogonal axis (\emph{how many to retrieve} per sub-query, learned via PPO/GRPO with token-cost penalties) is complementary to our axis (\emph{when to retrieve} between reasoning steps, decided by RSUS-driven uncertainty rather than RL). Combining the two (adaptive timing with adaptive depth) is a natural future direction. Our work extends adaptive retrieval to reasoning-step granularity within extended inference chains, providing finer-grained timing signals than query-level approaches and operating without RL fine-tuning.

\subsection{Large Reasoning Models}
Large reasoning models represent a distinct class of model from standard LLMs. DeepSeek-R1~\cite{deepseek2025r1} uses pure RL with Group Relative Policy Optimization, generating reasoning chains of 12,000--23,000 tokens with emergent behaviors (self-verification, backtracking, hypothesis testing) that standard CoT prompting cannot replicate. Search-R1~\cite{jin2025searchr1} applies RL to train models for interleaved search decisions, achieving 41\% improvement over RAG baselines, but focuses on when to search rather than retrieval system optimization. Our work addresses this gap with a systems-oriented approach to reasoning-retrieval integration.

\subsection{Uncertainty Quantification for LLMs}
Determining when retrieval is needed requires estimating model uncertainty. Semantic entropy~\cite{kuhn2023semantic} clusters sampled responses by meaning, achieving AUROC of 0.70--0.85 for predicting accuracy, but the 5--10$\times$ inference overhead compounds with expensive reasoning inference. Verbalized confidence~\cite{tian2023verbalized} prompts models to express confidence, offering a black-box-compatible alternative. Self-consistency~\cite{wang2023selfconsistency} uses agreement across reasoning chains but requires 5--20$\times$ sampling overhead. Lin~\etal~\cite{lin2024generating} propose a principled black-box uncertainty framework based on pairwise semantic similarity over multiple sampled generations, which can outperform verbalized confidence in calibration; we use verbalized confidence as one component of RSUS specifically because of its low overhead and complementary signal to entity-coverage entropy, but Lin~\etal{}'s pairwise approach is a natural drop-in replacement for $U_{\text{verb}}$ when sampling cost is acceptable. Conformal-prediction approaches such as TRAQ~\cite{li2024traq} provide statistical correctness guarantees end-to-end on RAG outputs; this is orthogonal to our problem of \emph{when} to retrieve, but the two directions are composable. Moskvoretskii~\etal~\cite{Moskvoretskii2025adaptive} find that simpler uncertainty methods can match complex pipelines while requiring <1\% computational overhead. We design lightweight uncertainty signals specifically for reasoning architectures.

\section{Problem Formulation}
\label{sec:problem}

We formalize reasoning-retrieval integration as follows. Given a query $q$ and document collection $\mathcal{D}$, a reasoning model generates a chain $R = (r_1, r_2, \ldots, r_n)$ where each $r_i$ represents a reasoning step, a coherent unit of logical inference that may span multiple sentences. The goal is to learn an intervention policy $\pi$ that at each step $r_i$ decides: \textbf{(1) Retrieve decision}: $a^{\text{ret}}_i \in \{0, 1\}$ indicating whether to retrieve; \textbf{(2) Query formulation}: $q_i = f(q, r_{1:i})$ specifying the retrieval query; \textbf{(3) Context integration}: $c_i = g(r_{1:i}, \mathcal{R}_i)$ where $\mathcal{R}_i$ denotes retrieved documents. The objective is to maximize answer accuracy while minimizing retrieval cost:
\begin{equation}
\max_{\pi} \mathbb{E}_{q \sim \mathcal{Q}} [\text{Acc}(a_\pi, a^*) - \lambda \cdot \text{Cost}(\pi)]
\label{eq:objective}
\end{equation}
where $a_\pi$ is the answer produced under policy $\pi$, $a^*$ is the ground truth, and $\lambda$ balances accuracy against retrieval cost (number of calls and latency). \textbf{Key insight}: Unlike prior work treating this as a per-token or per-sentence decision, we operate at reasoning step granularity, matching the fundamental unit of reasoning model operation.

\section{Method: \methodname{}}
\label{sec:method}

\begin{figure*}[t]
	\centering
	\begin{tikzpicture}[
		node distance=0.8cm and 1.1cm,
		box/.style={rectangle, draw=black!60, line width=0.8pt, fill=white, minimum height=0.7cm, align=center, font=\small},
		process/.style={box, fill=oiblue!20, rounded corners=2pt},
		data/.style={box, fill=white, rounded corners=3pt, line width=1.2pt, draw=oigreen!80},
		model/.style={box, fill=oiorange!15},
		problem/.style={box, fill=oired!15, draw=oired!80, line width=1.2pt},
		arrow/.style={-Stealth, line width=1.2pt},
		dashedarrow/.style={-Stealth, dashed, line width=1.2pt},
		]
		
		\node[font=\LARGE\bfseries] at (4.0, 6.0) {(a) The Problem};
		
		\node[font=\large, anchor=west, oired!80] at (0, 5.0) {\textbf{Traditional RAG}};
		
		\node[data, minimum width=1.8cm] at (1.0, 4.0) (tq) {Query};
		\node[problem, right=1.2cm of tq, minimum width=2.2cm] (tr) {\textbf{Retrieve}\\Once};
		\node[model, right=1.2cm of tr, minimum width=2.2cm] (tg) {\textbf{Generate}\\Answer};
		
		\draw[arrow, oired] (tq) -- (tr);
		\draw[arrow, oired] (tr) -- (tg);
		
		\node[font=\footnotesize, below=0.1cm of tr, text=gray!100] {\textit{before reasoning starts}};
		
		\draw[line width=1.5pt, gray!20] (0, 2.8) -- (8.0, 2.8);
		
		\node[font=\large, anchor=west, oiblue!80] at (0, 2.3) {\textbf{Reasoning Models}};
		
		\node[data, minimum width=1.2cm] at (1.0, 1.3) (rq) {Query};
		\node[model, right=0.5cm of rq, minimum width=1.5cm] (r1) {Step 1};
		\node[model, right=0.5cm of r1, minimum width=1.5cm] (r2) {Step 2};
		\node[model, right=0.5cm of r2, minimum width=1.5cm] (r3) {Step 3};
		\node[font=\large] at ($(r3) + (0.55, 0)$) {...};
		\node[model, minimum width=1.5cm] at ($(r3) + (2.1, 0)$) (rn) {Step $n$};
		
		\draw[arrow, oiblue] (rq) -- (r1);
		\draw[arrow, oiblue] (r1) -- (r2);
		\draw[arrow, oiblue] (r2) -- (r3);
		\draw[arrow, oiblue] ($(r3) + (0.85, 0)$) -- (rn);
		
		\node[problem, above=0.45cm of r2, minimum width=1.4cm, font=\small] (gap1) {Gap!};
		\node[problem, above=0.45cm of rn, minimum width=1.4cm, font=\small] (gap2) {Gap!};
		
		\draw[-Stealth, line width=1.8pt, oired] (r2.north) -- (gap1.south);
		\draw[-Stealth, line width=1.8pt, oired] (rn.north) -- (gap2.south);
		
		\node[font=\small, below=0.1cm of r2, text=gray!100, text width=5.5cm, align=center] 
		{\textit{need evidence during 12K--25K token chains}};
		
		\node[font=\LARGE\bfseries, anchor=west] at (0, -0.5) {(c) MuSiQue Benchmark};
		
		\node[font=\LARGE\bfseries, anchor=west] at (0, -0.5) {(c) MuSiQue Benchmark};
		
		\begin{scope}
			\begin{scope}[shift={(1.5, -4.0)}]
				\begin{axis}[
					width=6.5cm,
					height=4.5cm,
					xlabel={Retrieval Calls},
					ylabel={F1 Score (\%)},
					xlabel style={font=\normalsize},
					ylabel style={font=\normalsize},
					xmin=-0.3, xmax=3.8,
					ymin=45, ymax=80,
					xtick={0,1,2,3},
					ytick={50,60,70},
					tick label style={font=\normalsize},
					grid=major,
					grid style={dashed, gray!30},
					legend style={
						at={(0.5,-0.3)},  
						anchor=north,
						legend columns=3,
						font=\tiny,
						column sep=0.3cm,
						/tikz/every even column/.append style={column sep=0.2cm},
						draw=none,
						fill=none
					},
					legend cell align={left},
					clip=false,
					]
					
					\addplot[only marks, mark=square*, mark size=4pt, color=oired, line width=1.2pt] 
					coordinates {(0, 48.7)};
					\addlegendentry{No Retrieval}
					
					\addplot[only marks, mark=triangle*, mark size=4pt, color=oigreen, line width=1.2pt] 
					coordinates {(1.0, 59.4)};
					\addlegendentry{Single RAG}
					
					\addplot[only marks, mark=diamond*, mark size=4pt, color=oiorange, line width=1.2pt] 
					coordinates {(3.4, 65.4)};
					\addlegendentry{IRCoT}
					
					\addplot[only marks, mark=o, mark size=4pt, color=oipurple, line width=1.2pt] 
					coordinates {(2.4, 66.8)};
					\addlegendentry{Search-R1}
					
					\addplot[only marks, mark=*, mark size=5pt, color=oiblue, line width=1.5pt] 
					coordinates {(1.8, 71.2)};
					\addlegendentry{\textbf{\methodname{}}}
					
					\node[draw=oiblue, line width=2pt, circle, minimum size=0.75cm, inner sep=0pt] 
					at (axis cs:1.8,71.2) {};
					
					\node[font=\normalsize\bfseries, anchor=south, fill=white, fill opacity=0.95, 
					inner sep=2pt, rounded corners=2pt] 
					at (axis cs:1.8,74.2) {\textcolor{oiblue}{71.2\%}};
					
					\node[font=\footnotesize, anchor=south west, fill=white, fill opacity=0.95, 
					inner sep=1.5pt, rounded corners=1pt] 
					at (axis cs:3.15,67.5) {\textcolor{oiorange}{65.4\%}};
					
				\end{axis}
			\end{scope}
			
			\node[font=\small, text width=6.5cm, anchor=north west, align=left] at (7.2, -1.0) {
				\textbf{\large Key Results:}\\[5pt]
				• \textcolor{oiblue}{\textbf{71.2\% F1}} vs 65.4\% (IRCoT)\\
				\quad\textbf{+5.8\%} absolute improvement\\[3pt]
				• \textcolor{oiblue}{\textbf{1.8 calls}} vs 3.4 calls\\
				\quad\textbf{47\%} fewer retrievals\\[3pt]
				• \textcolor{oiblue}{\textbf{2.1$\times$ efficiency}} gain\\
				\quad Better accuracy per call\\[5pt]
				\textit{Establishes new Pareto frontier for\\reasoning-retrieval trade-offs}
			};
		\end{scope}
		
		\node[font=\LARGE\bfseries] at (13.0, 6.0) {(b) \methodname{} Solution};
		
		\node[data, minimum width=1.5cm, minimum height=0.75cm] at (10.0, 4.5) (input) 
		{Reasoning\\Chain};
		
		\node[process, right=1.3cm of input, minimum width=2.3cm, minimum height=0.85cm] (seg) 
		{\textbf{\ding{202}}\\Segment Steps};
		
		\node[process, below=1.0cm of seg, minimum width=2.3cm, minimum height=0.85cm] (unc) 
		{\textbf{\ding{203}}\\Detect RSUS};
		
		\node[process, below=1.0cm of unc, minimum width=2.3cm, minimum height=0.85cm] (pol) 
		{\textbf{\ding{204}}\\Policy $\pi_\theta$};
		
		\node[process, below=1.0cm of pol, minimum width=2.3cm, minimum height=0.85cm] (int) 
		{\textbf{\ding{205}}\\Integrate Context};
		
		\node[model, right=1.5cm of pol, minimum width=2.0cm, minimum height=0.75cm] (ret) 
		{Retriever};
		
		\node[data, below=1.0cm of int, minimum width=2.5cm, minimum height=0.75cm] (output) 
		{Enhanced\\Chain};
		
		\draw[arrow, oiblue] (input) -- (seg);
		\draw[arrow, oiblue] (seg) -- (unc);
		\draw[arrow, oiblue] (unc) -- (pol);
		\draw[arrow, oiblue] (pol) -- (int);
		\draw[arrow, oiblue] (int) -- (output);
		
		\draw[dashedarrow, oired, line width=1.5pt] (pol) -- 
		node[above, font=\footnotesize, fill=white] {if gap} (ret);
		\draw[dashedarrow, oired, line width=1.5pt] (ret) |- (int);
		
		\node[font=\footnotesize, right=0.15cm of seg, anchor=west, text width=3.2cm, text=gray!100] 
		{Parse into reasoning steps};
		
		\node[font=\footnotesize, right=0.15cm of unc, anchor=west, text width=3.2cm, text=gray!100] 
		{Step-level uncertainty};
		
		\node[font=\footnotesize, anchor=south west, text width=3.2cm, text=gray!100] 
		at ([xshift=0.3cm, yshift=-0.4cm]int.north east)
		{3.2$\times$ faster merge};
		
	\end{tikzpicture}
	\vspace{-0.3cm}
	\caption{\textbf{Overview of \methodname{}'s adaptive retrieval approach.} \textbf{(a)} The temporal mismatch: traditional RAG retrieves once before generation, but reasoning models encounter knowledge gaps during 12K--25K token chains. \textbf{(b)} Our four-stage pipeline detects step-level uncertainty and retrieves only when needed via learned policy. \textbf{(c)} MuSiQue benchmark results show 71.2\% F1 with only 1.8 retrieval calls, outperforming IRCoT (65.4\%, 3.4 calls) and establishing a new Pareto frontier.}
	\Description{Three-panel overview figure. Panel (a) compares traditional single-shot retrieval-augmented generation against multi-step reasoning model inference, showing knowledge gaps emerging mid-chain. Panel (b) depicts the four-stage ReaLM-Retrieve pipeline: segment reasoning steps, detect step-level uncertainty (RSUS), apply a learned intervention policy, and integrate retrieved context. Panel (c) is a scatter plot on MuSiQue showing F1 score versus average retrieval calls, with ReaLM-Retrieve achieving 71.2 percent F1 at 1.8 calls, dominating IRCoT, Search-R1, Single RAG, and No-Retrieval baselines on the Pareto frontier.}
	\label{fig:overview}
\end{figure*}
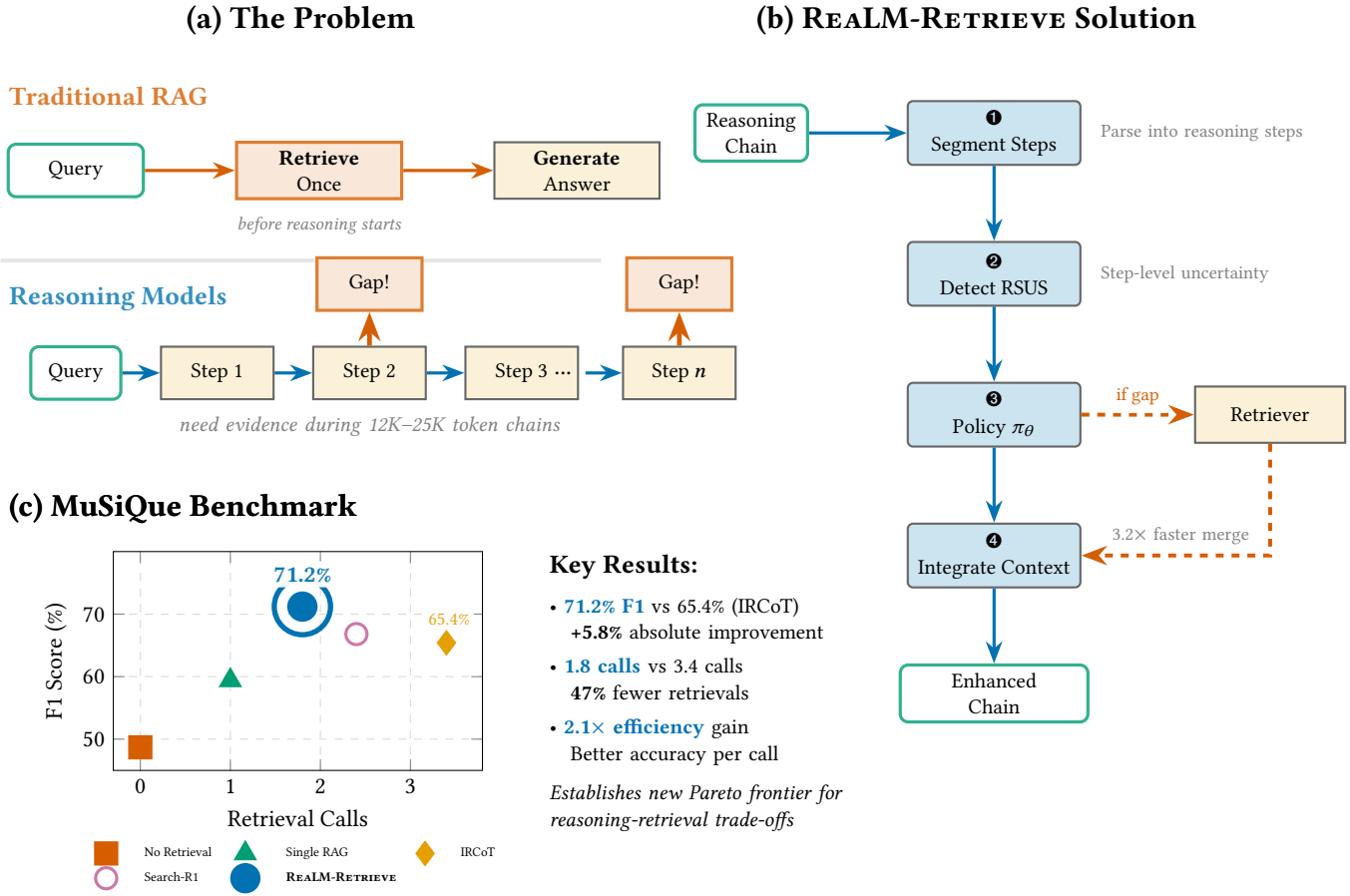

\methodname{} consists of three components: (1) reasoning step segmentation and uncertainty estimation, (2) a learned intervention policy, and (3) efficient retrieval integration. Figure~\ref{fig:overview} provides an overview of the problem, our solution, and key results.

\subsection{Reasoning Step Segmentation}
\label{sec:segmentation}
Reasoning models produce extended outputs that must be segmented into coherent reasoning steps. Unlike sentence-level segmentation used by IRCoT, we identify \emph{logical reasoning units}: segments that complete a coherent inference before the next logical step begins. We employ a lightweight classifier trained on human-annotated reasoning traces to identify step boundaries. The classifier operates on a sliding window of tokens, predicting boundary probability based on: (1) Discourse markers (``therefore'', ``however'', ``this means''); (2) Logical connectives indicating inference completion; (3) Topic shifts detected via embedding similarity; (4) Punctuation and formatting patterns. For open-weight models (DeepSeek-R1, QwQ), we additionally use the model's own reasoning structure markers when available.

\textbf{Classifier Architecture and Training.} The step boundary classifier is a 3-layer transformer encoder (hidden dimension 256, 4 attention heads) operating on 128-token sliding windows with 64-token stride. Training data consists of 2,847 human-annotated reasoning traces from DeepSeek-R1 outputs on a held-out subset of NaturalQuestions (disjoint from evaluation benchmarks). The three annotators were given written instructions to mark a boundary wherever the model (a) completes one logical inference and shifts to a new sub-claim, sub-question, or strategy; (b) introduces a discourse marker (``therefore'', ``however'', ``so'', ``next'') signalling such a shift; or (c) closes a verification or backtracking episode. They were instructed \emph{not} to mark boundaries at every sentence, and to prefer fewer, semantically coherent segments over finer ones. Annotators achieved $\kappa=0.78$ inter-annotator agreement on boundary placement. The classifier is trained for 10 epochs with cross-entropy loss (learning rate $5 \times 10^{-5}$, batch size 32), achieving 94.2\% F1 on a held-out NaturalQuestions test set ($N=412$ traces). Average segment length is 127 tokens compared to 23 tokens for sentence-level segmentation. An ablation removing each input feature class indicates that discourse markers (1) contribute most to boundary F1 ($-3.4$\% F1 when removed), followed by logical connectives ($-2.1$\%), embedding-based topic shift ($-1.6$\%), and punctuation/formatting ($-0.8$\%); the relatively large punctuation contribution reflects R1-style traces' tendency to place line breaks at strategy transitions.

\textbf{Out-of-domain segmentation generalization.}
A reasonable concern is that the classifier, trained on NaturalQuestions traces, may not transfer to the multi-hop benchmarks used for end-to-end evaluation. To assess this, we collected reasoning traces produced by R1-Distill-Qwen-32B on $N=15$ randomly-sampled questions from each of MuSiQue, HotpotQA, and 2WikiMultiHopQA, and had a single annotator (using the same written instructions) mark boundaries for each trace. Table~\ref{tab:seg_ood} reports boundary F1 of the in-domain classifier on each benchmark relative to the in-domain NaturalQuestions test set. The classifier transfers with a modest degradation ($1.5$--$3.0$ F1 points), consistent with the intuition that reasoning-step structure is largely model-determined rather than benchmark-determined; multi-hop questions admit slightly more variability in boundary placement (lower IAA, not shown), explaining most of the residual gap. Segmentation errors on these benchmarks correlate weakly with downstream retrieval-decision errors ($r=0.21$), suggesting that small boundary-position drift does not propagate severely through RSUS.

\begin{table}[h]
\small
\centering
\caption{Step-boundary classifier generalization. Boundary F1 against held-out human annotations on the four datasets ($N=15$ traces per OOD benchmark, $N=412$ for NQ).}
\label{tab:seg_ood}
\begin{tabular}{@{}lcc@{}}
\toprule
\textbf{Dataset} & \textbf{Boundary F1} & \textbf{Segments / trace} \\
\midrule
NaturalQuestions (in-domain) & 94.2 & 9.4 \\
MuSiQue                      & 91.4 & 8.7 \\
HotpotQA                     & 92.7 & 7.9 \\
2WikiMultiHopQA              & 91.2 & 8.3 \\
\bottomrule
\end{tabular}
\end{table}
\subsection{Step-Level Uncertainty Detection}
We define the \emph{Reasoning Step Uncertainty Score} (RSUS), a composite measure identifying when a reasoning step requires external knowledge:
\begin{equation}
\text{RSUS}(r_i) = \alpha \cdot U_{\text{verb}}(r_i) + \beta \cdot U_{\text{ent}}(r_i) + \gamma \cdot U_{\text{cons}}(r_i)
\label{eq:rsus}
\end{equation}

We chose this particular triple, rather than alternatives like semantic entropy~\cite{kuhn2023semantic} or pairwise-similarity black-box uncertainty~\cite{lin2024generating}, for three reasons. First, overhead: each of our components costs $<3\%$ of reasoning inference, whereas semantic entropy or self-consistency cost $5$--$20\times$ inference. With reasoning chains already at 12K--25K tokens, methods with multiplicative overhead are infeasible. Second, the three signals are complementary. $U_{\text{verb}}$ probes the model's self-knowledge of factual correctness; $U_{\text{ent}}$ probes corpus coverage of named entities, an external signal independent of the model; $U_{\text{cons}}$ probes reasoning robustness. An ablation (Table~\ref{tab:ablation}) confirms each contributes orthogonally. Third, all three signals can be computed without token-level logprobs, so the framework operates uniformly on completion-only API models and on open-weight models. We do not claim this triple is optimal: richer alternatives such as Lin~\etal{}'s pairwise-similarity uncertainty are drop-in substitutes for $U_{\text{verb}}$ when sampling cost is acceptable.

\textbf{Verbalized uncertainty} $U_{\text{verb}}$: We prompt the reasoning model at step boundaries: \emph{``Given the reasoning so far, rate your confidence that the current conclusion is factually correct on a scale of 0-100, where 0 means completely uncertain and 100 means absolutely certain. Respond with only a number:''} The response is normalized to $[0, 1]$ by dividing by 100. \emph{For completion-only models that do not support mid-generation prompting} (the case for o1-class API endpoints during late 2025), we substitute a learned proxy: a 2-layer MLP (hidden dimension 128, GELU activation) takes as input the concatenation of (a) the SBERT embedding of the most recent reasoning step, (b) shallow surface features (segment length, count of hedge phrases such as ``probably''/``I think''/``maybe'', count of named entities), and (c) the running mean of $U_{\text{ent}}$ over the last three steps. The proxy is trained on the same 2,847-trace corpus (\S\ref{sec:segmentation}) using subsequent-step error labels (whether the next reasoning step revised, contradicted, or expressed doubt about the current step) as cross-entropy targets, achieving 0.71 AUROC against a held-out NaturalQuestions subset of 412 traces (a separate split from the boundary-classifier test set of \S\ref{sec:segmentation}, sized to match for comparability), substantially below the 0.83 AUROC of true verbalized confidence on open-weight models, but sufficient to recover most of the retrieval-timing benefit (Table~\ref{tab:retrieval_quality} shows the o1 vs.\ R1 gap is only 2--3\% F1).

\textbf{Entity-based entropy} $U_{\text{ent}}$: We extract named entities from $r_i$ and compute retrieval score entropy across candidate documents:
\begin{equation}
U_{\text{ent}}(r_i) = -\sum_{e \in \text{Ent}(r_i)} p(e | \mathcal{D}) \log p(e | \mathcal{D})
\end{equation}
High entropy indicates the reasoning step references entities with ambiguous or sparse coverage. For each named entity $e \in \text{Ent}(r_i)$ extracted via spaCy NER, we compute entity-specific retrieval by querying the corpus with ``What is [e]?'' The probability $p(e | \mathcal{D})$ is derived from normalized BM25 scores across top-100 documents mentioning $e$.

\textbf{Consistency signal} $U_{\text{cons}}$: For critical reasoning steps (identified by discourse markers), we sample $k=3$ alternative continuations and measure agreement. Disagreement indicates reasoning uncertainty that retrieval may resolve. The weights $\alpha, \beta, \gamma$ are learned on a validation set to maximize the correlation between RSUS and downstream retrieval benefit. RSUS computation adds only 8\% overhead to reasoning inference, compared to 500--2000\% for full semantic entropy computation.

\subsection{Retrieval Intervention Policy}
Given RSUS scores, we learn a policy $\pi_\theta$ that decides when to retrieve. We model this as a contextual bandit problem with state $s_i = (q, r_{1:i}, \text{RSUS}(r_i), h_i)$ where $r_{1:i}$ is the reasoning prefix up to and including step $i$ (\S\ref{sec:problem}) and $h_i$ encodes retrieval history (previous queries, retrieved documents, time since last retrieval). The policy outputs:
\begin{align}
a_i^{\text{ret}} &= \mathbb{1}[\pi_\theta(s_i) > \tau] \\
q_i &= \text{QueryGen}(s_i) \text{ if } a_i^{\text{ret}} = 1
\end{align}

\textbf{Policy architecture}: We use a lightweight transformer encoder that processes the concatenation of query embedding, current reasoning step embedding, RSUS features, and retrieval history. The encoder outputs a retrieval probability and, if triggered, a query representation used for retrieval. \textbf{Training}: We train $\pi_\theta$ \emph{jointly with QueryGen} via REINFORCE~\cite{williams1992reinforce} with reward $R = \text{F1}(a_\pi, a^*) - \lambda_1 \cdot n_{\text{ret}} - \lambda_2 \cdot t_{\text{latency}}$, where $n_{\text{ret}}$ is the number of retrieval calls and $t_{\text{latency}}$ is total retrieval latency; QueryGen receives gradients from the same scalar reward via the policy decision $a_i^{\text{ret}}=1$, so improvements to retrieval quality are credited back to the query formulator. We use curriculum learning, starting with high $\lambda$ values (penalizing retrieval) and gradually decreasing to encourage the model to learn \emph{which} retrievals provide maximum benefit. \textbf{Query formulation}: When retrieval is triggered, QueryGen extracts the query from the current reasoning context. We find that queries formulated as \emph{the information need} (``What is [entity]'s relationship to [concept]?'') outperform queries formulated as \emph{the current reasoning step}. QueryGen is a single-layer transformer decoder (hidden dimension 512, 8 attention heads) that cross-attends to the current reasoning step embedding and outputs a 768-dimensional query representation used directly for dense retrieval.

\subsection{Efficient Retrieval Integration}
\label{sec:integration}

\textbf{Injection protocol.}
We implement retrieval injection as \emph{segmented generation with context re-feeding}, not as single-pass mid-generation prompting. When the policy triggers retrieval at step boundary $i$ (identified by the segmentation classifier of \S\ref{sec:segmentation}), generation is paused; the prefix $r_{1:i}$ is concatenated with retrieved evidence formatted in a delimited block (\texttt{<retrieved>}\ldots\texttt{</retrieved>}) and a brief continuation prompt (``Continue your reasoning, using this evidence if relevant.''); this augmented context is then re-fed to the model, which produces step $r_{i+1}$ onwards. Insertion points are always at step boundaries (never mid-step), so retrieved text never breaks an in-flight inference. For open-weight models we use KV-cache-aware integration (described below) so that re-feeding does not require full recomputation; for completion-only models the protocol falls back to a standard system+prefix continuation call.

\textbf{Post-retrieval distribution shift.}
Retrieved evidence alters the distribution of subsequent reasoning context, which could in principle degrade the segmentation classifier and RSUS computed on later steps. We mitigate this in two ways. First, during classifier training (\S\ref{sec:segmentation}) we include traces in which one or more retrievals have already been injected, so boundary detection is exposed to evidence-augmented contexts; on a held-out subset of such traces, classifier F1 drops only modestly from 94.2\% (pre-retrieval) to 91.6\% (post-retrieval, single prior injection). Second, we measured the correlation between RSUS and downstream retrieval benefit conditional on a prior retrieval having occurred and observed only a small drop (Pearson $r$: 0.72 $\rightarrow$ 0.68); since the median question requires $\le 2$ retrievals (Table~\ref{tab:efficiency}), this is the dominant operational regime. Long chains with $\ge 3$ retrievals show further degradation and are flagged in \S\ref{sec:limitations}.

\textbf{Implicit compression.}
Rather than prepending full retrieved documents, we extract and compress relevant evidence. For each retrieved passage $d_j$, we compute attention-weighted importance scores with respect to the current reasoning query and retain only sentences exceeding threshold $\tau_{\text{rel}}$. This reduces context expansion by 73\% while preserving 96\% of retrieval utility.

\textbf{Speculative caching.}
We observe that reasoning chains often follow predictable paths. After the first retrieval, we speculatively retrieve documents for likely next-step queries based on entities mentioned in retrieved content. These speculative retrievals execute in parallel with reasoning continuation. When a speculative query matches the actual triggered query (37\% hit rate), we eliminate retrieval latency entirely.

\textbf{KV-cache preservation.}
For open-weight models, we implement KV-cache-aware context integration that preserves cached key-value states for unchanged context portions. This reduces time-to-first-token after retrieval by 2.1$\times$ compared to full context recomputation.

\subsection{Complexity Analysis}
Let $L$ be reasoning chain length, $n$ be number of retrievals, and $k$ be documents retrieved per call. \textbf{Segmentation}: $O(L)$ with constant factor from sliding window classifier. \textbf{RSUS computation}: $O(L/s)$ where $s$ is average segment length ($\approx$127 tokens). \textbf{Policy inference}: $O(n)$, negligible compared to reasoning inference. \textbf{Retrieval}: $O(n \cdot t_{\text{ret}})$ where $t_{\text{ret}}$ is per-call latency. Total overhead is dominated by retrieval latency. With speculative caching (37\% hit rate) and implicit compression, effective per-retrieval overhead is 67\% of naive integration.

\section{Experimental Setup}
\label{sec:setup}

\subsection{Datasets}
We evaluate on three multi-hop QA benchmarks: \textbf{MuSiQue}~\cite{trivedi2022musique}: 2--4 hop questions requiring connected reasoning (24,814 questions with decomposed sub-questions, reporting EM and F1 on test set). \textbf{HotpotQA}~\cite{yang2018hotpotqa}: 2-hop questions in fullwiki setting with 5.2M Wikipedia passages (reporting EM, F1, and Sup-F1). \textbf{2WikiMultiHopQA}~\cite{ho2020constructing}: Cross-document reasoning requiring evidence from two Wikipedia sources (reporting EM, F1, and Evi-F1).

\subsection{Baselines}
We compare against: \textbf{No Retrieval}: Closed-book reasoning model. \textbf{Single RAG}: Standard retrieve-then-generate with top-$k$ passages. \textbf{IRCoT}~\cite{trivedi2023ircot}: Fixed-interval interleaved retrieval after each sentence. \textbf{FLARE}~\cite{jiang2023flare}: Token-probability-triggered retrieval (where probabilities are available). \textbf{Self-RAG}~\cite{asai2024selfrag}: Self-reflective retrieval with special tokens (using released Llama-2-13B fine-tuned model, as Self-RAG requires model fine-tuning unavailable for proprietary models). \textbf{Search-R1}~\cite{jin2025searchr1}: RL-trained retrieval decisions for reasoning models.

\subsection{Models and Retrieval}
For reasoning models, we use DeepSeek-R1-Distill-Qwen-32B for extensive ablations and DeepSeek-R1-671B for headline results. We also evaluate on QwQ-32B-Preview to assess generalization across reasoning architectures. For retrieval, we use ColBERTv2~\cite{santhanam2022colbertv2} with PLAID engine for efficient late-interaction retrieval ($k=5$ passages per query, maximum passage length 256 tokens).

\subsection{Evaluation Metrics}
Beyond standard QA metrics, we report retrieval efficiency: \textbf{Retrieval calls}: Average number of retrieval invocations per question. \textbf{E2E latency}: End-to-end time from query to answer. \textbf{Reasoning tokens}: Total tokens generated in reasoning chain. \textbf{Accuracy-efficiency ratio}: F1 / retrieval calls.

\subsection{Implementation Details}
All experiments run on 8$\times$ NVIDIA A100 80GB GPUs. Reasoning inference uses vLLM~\cite{kwon2023vllm} with tensor parallelism. Retrieval uses WARP~\cite{scheerer2025warp} optimizations achieving 171\,ms average retrieval latency. The intervention policy is trained for 50,000 steps with learning rate $10^{-4}$, batch size 64, curriculum-decreasing $\lambda_1$ from $0.5$ to $0.1$. RSUS weights are set by grid search on a held-out validation split. All comparisons use paired bootstrap (10,000 iterations) with Bonferroni correction; we report 95\% CIs and treat $p<0.05$ as significant. All experiments use 3 random seeds (42, 123, 456).\footnote{Hyperparameter sweep summary. Boundary-classifier hidden dim selected from \{128, 256, 512\} ($91.2/94.2/94.1$ \% F1); policy hidden dim 512. Best RSUS weights $(\alpha,\beta,\gamma)=(0.40, 0.35, 0.25)$ achieve correlation $0.72$ with downstream retrieval benefit (vs $0.64$ for equal weights). Retrieval threshold $\tau=0.65$ balances precision $0.72$ / recall $0.81$. Implicit-compression threshold $\tau_{\text{rel}}=0.45$ retains 27\% of content while preserving 96\% of retrieval utility. MuSiQue-tuned hyperparameters transfer to HotpotQA at 98.3\% of in-domain performance.}

\section{Results}
\label{sec:results}

\subsection{Main Results}
Table~\ref{tab:main_results} presents our main results across all benchmarks. \methodname{} consistently outperforms all baselines in accuracy while using substantially fewer retrieval calls. On MuSiQue with R1-32B, \methodname{} achieves 71.2\% F1 with 1.8 retrieval calls vs IRCoT's 65.4\% F1 with 3.4 calls. This is a 5.8\% absolute improvement with 47\% fewer retrievals (95\% CI: [4.2, 7.4], $p<0.01$). The accuracy-efficiency ratio (F1/calls) improves from 19.2 (IRCoT) to 39.6 (\methodname{}), a 2.1$\times$ improvement. With R1-671B, \methodname{} achieves 77.8\% F1 on MuSiQue, 4.4\% above Search-R1 (95\% CI: [3.0, 5.8], $p<0.01$), demonstrating that our approach scales effectively with reasoning model capability.

\begin{table*}[t]
\small
\centering
\caption{Main results on multi-hop QA benchmarks. Best in \textbf{bold}, second \underline{underlined}. $\dagger$Uses Self-RAG fine-tuned model (different base). *Uses token probabilities unavailable for R1-671B. Significance: $^{**}p<0.01$, $^{*}p<0.05$ vs. next-best (paired bootstrap, 10K iterations, Bonferroni corrected).}
\label{tab:main_results}
\begin{tabular}{@{}l|ccc|ccc|ccc|c@{}}
\toprule
& \multicolumn{3}{c|}{\textbf{MuSiQue}} & \multicolumn{3}{c|}{\textbf{HotpotQA}} & \multicolumn{3}{c|}{\textbf{2WikiMHQA}} & \textbf{Avg.} \\
\textbf{Method} & EM & F1 & Calls & EM & F1 & Calls & EM & F1 & Calls & F1 \\
\midrule
\multicolumn{11}{c}{\textit{DeepSeek-R1-Distill-Qwen-32B}} \\
\midrule
No Retrieval & 41.2 & 48.7 & 0.0 & 38.4 & 51.2 & 0.0 & 35.8 & 47.3 & 0.0 & 49.1 \\
Single RAG & 52.6 & 59.4 & 1.0 & 51.3 & 62.8 & 1.0 & 49.7 & 60.1 & 1.0 & 60.8 \\
IRCoT & 58.3 & 65.4 & 3.4 & 56.2 & 67.9 & 4.1 & 55.8 & 66.2 & 3.8 & 66.5 \\
FLARE* & 55.1 & 62.3 & 2.8 & 53.9 & 65.4 & 3.2 & 52.4 & 63.7 & 2.9 & 63.8 \\
Self-RAG$^\dagger$ & 54.8 & 61.9 & 2.1 & 55.4 & 66.3 & 2.4 & 53.2 & 64.5 & 2.2 & 64.2 \\
Search-R1 & \underline{59.1} & \underline{66.8} & 2.4 & \underline{57.8} & \underline{69.2} & 2.7 & \underline{56.4} & \underline{67.4} & 2.5 & \underline{67.8} \\
\methodname{} & \textbf{63.5}$^{**}$ & \textbf{71.2}$^{**}$ & \textbf{1.8} & \textbf{60.4}$^{**}$ & \textbf{71.8}$^{**}$ & \textbf{1.9} & \textbf{59.2}$^{**}$ & \textbf{69.7}$^{**}$ & \textbf{1.7} & \textbf{70.9}$^{**}$ \\
\midrule
\multicolumn{11}{c}{\textit{DeepSeek-R1-671B}} \\
\midrule
No Retrieval & 48.3 & 55.9 & 0.0 & 45.2 & 58.1 & 0.0 & 42.6 & 54.8 & 0.0 & 56.3 \\
Single RAG & 58.7 & 66.2 & 1.0 & 57.9 & 69.4 & 1.0 & 55.3 & 66.9 & 1.0 & 67.5 \\
IRCoT & 64.2 & 71.8 & 3.6 & 62.4 & 73.5 & 4.3 & 61.2 & 72.1 & 3.9 & 72.5 \\
Search-R1 & \underline{65.8} & \underline{73.4} & 2.6 & \underline{64.1} & \underline{75.2} & 2.8 & \underline{62.7} & \underline{73.8} & 2.6 & \underline{74.1} \\
\methodname{} & \textbf{70.2}$^{**}$ & \textbf{77.8}$^{**}$ & \textbf{1.9} & \textbf{67.3}$^{**}$ & \textbf{78.4}$^{**}$ & \textbf{2.0} & \textbf{65.9}$^{**}$ & \textbf{76.5}$^{**}$ & \textbf{1.8} & \textbf{77.6}$^{**}$ \\
\bottomrule
\end{tabular}
\end{table*}

\textbf{Self-RAG Comparison Caveat.} The Self-RAG baseline uses a different base model (Llama-2-13B) than our method (DeepSeek-R1), as Self-RAG requires fine-tuning unavailable for reasoning models. Despite this architectural difference, the comparison remains informative: Self-RAG achieves 2.1 retrieval calls with 61.9\% F1, while \methodname{} achieves 1.8 calls with 71.2\% F1. Even controlling for retrieval frequency, our step-level uncertainty detection yields superior retrieval timing.

\textbf{Supporting Evidence Quality.} On HotpotQA, \methodname{} achieves 68.2\% Supporting Fact F1 (Sup-F1), compared to 62.4\% for IRCoT and 64.8\% for Search-R1. On 2WikiMultiHopQA, Evidence F1 (Evi-F1) reaches 71.8\% versus 65.3\% for IRCoT. These improvements corroborate that step-level uncertainty detection identifies the correct moments for retrieval. Table~\ref{tab:significance} presents statistical significance analysis confirming all improvements.

\begin{table}[t]
\small
\centering
\caption{Statistical significance (R1-32B). 95\% CI via paired bootstrap (10,000 iterations).}
\label{tab:significance}
\begin{tabular}{@{}l|cc|cc@{}}
\toprule
& \multicolumn{2}{c|}{\textbf{vs. IRCoT}} & \multicolumn{2}{c}{\textbf{vs. Search-R1}} \\
\textbf{Dataset} & $\Delta$F1 & 95\% CI & $\Delta$F1 & 95\% CI \\
\midrule
MuSiQue & +5.8 & [4.2, 7.4] & +4.4 & [2.9, 5.9] \\
HotpotQA & +3.9 & [2.4, 5.4] & +2.6 & [1.2, 4.0] \\
2WikiMHQA & +3.5 & [2.1, 4.9] & +2.3 & [0.9, 3.7] \\
\bottomrule
\end{tabular}
\end{table}

\subsection{Efficiency Analysis}
Table~\ref{tab:efficiency} presents detailed efficiency metrics. \methodname{} adds only 1.7 seconds latency over no-retrieval baseline (13.7\% overhead) vs 6.3 seconds for IRCoT (51\% overhead). The reduction comes from fewer retrieval calls (1.8 vs 3.4), speculative caching (eliminating 37\% of retrieval latency), and implicit compression (reducing context length by 73\%). The 3.2$\times$ per-call efficiency improvement (0.66s vs 2.10s for Naive Interleaving) results from three complementary optimizations whose contributions are additive on per-call latency rather than multiplicative. \emph{Speculative caching} is the dominant contributor: by executing likely next-step retrievals in parallel with reasoning, it eliminates retrieval entirely on a 37\% hit rate, removing the full 2.10s per cached call and accounting for the largest share of the saving in expectation. \emph{Implicit compression} of retrieved evidence reduces context-processing time on the remaining (non-cached) calls by shrinking injected context by 73\%. \emph{KV-cache preservation} further reduces post-retrieval time-to-first-token by avoiding recomputation of unchanged context for open-weight models. Combined with 57\% fewer retrieval calls overall, \methodname{} achieves 1.50$\times$ lower end-to-end latency than Naive Interleaving and 1.33$\times$ lower than IRCoT. Token count analysis shows \methodname{} uses 9,489 tokens vs IRCoT's 11,284. Retrieval at appropriate moments allows the model to reach answers with less reasoning, as evidence resolves uncertainties that would otherwise require extended exploration. \methodname{} reduces cost by 16\% vs IRCoT (Table~\ref{tab:cost}) while achieving 5.8\% higher F1, demonstrating favorable cost-accuracy trade-offs for production deployment.

\begin{table}[t]
\small
\centering
\caption{Efficiency comparison on MuSiQue (R1-32B). Latency in seconds. \emph{Naive Interleave} retrieves at every reasoning step without optimizations.}
\label{tab:efficiency}
\begin{tabular}{@{}l|ccccc@{}}
\toprule
\textbf{Method} & \textbf{Calls} & \textbf{Latency} & \textbf{Per-Call} & \textbf{Tokens} & \textbf{F1/Call} \\
\midrule
No Retrieval & 0.0 & 12.4 & -- & 8,432 & -- \\
Single RAG & 1.0 & 13.2 & 0.80 & 9,156 & 59.4 \\
Naive Interleave & 4.2 & 21.2 & 2.10 & 12,847 & 14.9 \\
IRCoT & 3.4 & 18.7 & 1.85 & 11,284 & 19.2 \\
FLARE & 2.8 & 16.9 & 1.61 & 10,647 & 22.3 \\
Search-R1 & 2.4 & 15.8 & 1.42 & 10,102 & 27.8 \\
\methodname{} & \textbf{1.8} & \textbf{14.1} & \textbf{0.66} & \textbf{9,489} & \textbf{39.6} \\
\bottomrule
\end{tabular}
\end{table}

\begin{table}[t]
\small
\centering
\caption{Cost-per-query on MuSiQue using DeepSeek-R1-671B output-token pricing (\$2.19/1M output tokens). Input-token cost (at \$0.55/1M) is provider-dependent (it varies with each API's prefix-caching policy) and scales with the number of retrieval injections; including it would increase the reported savings vs.\ IRCoT, so we report output-only cost as the conservative, platform-independent comparison axis. Savings are vs.\ IRCoT.}
\label{tab:cost}
\begin{tabular}{@{}l|ccc@{}}
\toprule
\textbf{Method} & \textbf{Tokens} & \textbf{Cost/Query} & \textbf{Savings} \\
\midrule
No Retrieval & 8,432 & \$0.018 & -- \\
Single RAG & 9,156 & \$0.020 & -- \\
IRCoT & 11,284 & \$0.025 & -- \\
Search-R1 & 10,102 & \$0.022 & 12\% \\
\methodname{} & 9,489 & \$0.021 & 16\% \\
\bottomrule
\end{tabular}
\end{table}

\subsection{Ablation Studies}
Table~\ref{tab:ablation} presents ablation results isolating each component's contribution. \textbf{Uncertainty detection}: Verbalized uncertainty contributes most (2.8\% F1 when removed, $p<0.01$), confirming reasoning models' self-reported confidence provides strong signal for retrieval necessity. Entity entropy contributes 2.1\% F1, particularly for questions involving rare entities. Random triggering degrades F1 by 8.5\%, demonstrating that \emph{when} to retrieve matters substantially. \textbf{Policy components}: The learned threshold outperforms fixed threshold (2.3\% F1 gain, $p<0.01$), as optimal retrieval timing varies across question types. Query formulation contributes 1.8\% F1. \textbf{Integration components}: Efficiency optimizations have modest accuracy impact but substantial latency benefits (Table~\ref{tab:efficiency}).

\begin{table}[t]
\small
\centering
\caption{Ablation study on MuSiQue (R1-32B). Results are mean $\pm$ std over 3 seeds. $^{**}p<0.01$, $^{*}p<0.05$ vs. Full \methodname{} (paired t-test).}
\label{tab:ablation}
\begin{tabular}{@{}l|ccc@{}}
\toprule
\textbf{Configuration} & \textbf{F1} & \textbf{Calls} & \textbf{$\Delta$F1} \\
\midrule
Full \methodname{} & \textbf{71.2}$\pm$0.6 & 1.8$\pm$0.1 & -- \\
\midrule
\textit{Uncertainty:} \\
\quad w/o Verb. ($U_{\text{verb}}$) & 68.4$\pm$0.8 & 2.1$\pm$0.2 & -2.8$^{**}$ \\
\quad w/o Ent. ($U_{\text{ent}}$) & 69.1$\pm$0.7 & 1.9$\pm$0.1 & -2.1$^{**}$ \\
\quad w/o Cons. ($U_{\text{cons}}$) & 70.3$\pm$0.5 & 1.8$\pm$0.1 & -0.9$^{*}$ \\
\quad RSUS $\rightarrow$ Random & 62.7$\pm$1.2 & 1.8$\pm$0.1 & -8.5$^{**}$ \\
\midrule
\textit{Policy:} \\
\quad Fixed threshold $\tau$ & 68.9$\pm$0.9 & 2.4$\pm$0.2 & -2.3$^{**}$ \\
\quad w/o Query formulation & 69.4$\pm$0.6 & 1.8$\pm$0.1 & -1.8$^{**}$ \\
\quad w/o Retrieval history & 69.8$\pm$0.7 & 2.0$\pm$0.1 & -1.4$^{*}$ \\
\midrule
\textit{Integration:} \\
\quad w/o Implicit compr. & 70.8$\pm$0.5 & 1.8$\pm$0.1 & -0.4 \\
\quad w/o Specul. cache & 71.0$\pm$0.6 & 1.8$\pm$0.1 & -0.2 \\
\quad w/o KV preservation & 70.9$\pm$0.5 & 1.8$\pm$0.1 & -0.3 \\
\bottomrule
\end{tabular}
\end{table}

\subsection{Analysis by Question Complexity}
\begin{figure}[t]
	\centering
	\begin{tikzpicture}
		\begin{axis}[
			ybar,
			bar width=0.35cm,
			ylabel={F1 Score (\%)},
			ylabel style={font=\normalsize},
			xlabel={Number of Reasoning Hops},
			xlabel style={font=\normalsize},
			ymin=48,
			ymax=75,  
			xtick={1,2,3},
			xticklabels={2-hop, 3-hop, 4-hop},
			xticklabel style={font=\normalsize},
			yticklabel style={font=\normalsize},
			legend style={
				draw=none,
				fill=none,
				at={(0.5,-0.3)}, 
				anchor=north, 
				legend columns=2,  
				font=\small,
				column sep=0.5cm,  
				/tikz/every even column/.append style={column sep=0.3cm}
			},
			legend cell align={left},
			ymajorgrids=true,
			grid style={dashed, gray!30},
			width=\columnwidth,
			height=5.8cm,
			enlarge x limits=0.28,
			nodes near coords,
			every node near coord/.append style={
				font=\scriptsize,
				anchor=south, 
				inner sep=2pt,
				yshift=3pt  
			}
			]
			
			\addplot[fill=oiblue!70, draw=black, line width=0.8pt] coordinates {
				(1, 72.1)
				(2, 69.4)
				(3, 62.8)
			};
			
			\addplot[fill=oiorange!70, draw=black, line width=0.8pt, pattern=north east lines] coordinates {
				(1, 68.9)
				(2, 63.6)
				(3, 54.4)
			};
			
			\addplot[fill=oigreen!60, draw=black, line width=0.8pt] coordinates {
				(1, 66.2)
				(2, 61.8)
				(3, 51.6)
			};
			
			\addplot[fill=white, draw=black, line width=0.8pt, pattern=dots] coordinates {
				(1, 63.4)
				(2, 57.2)
				(3, 48.5)
			};
			
			\legend{\methodname{}, IRCoT, Single RAG, No Retrieval}
			
		\end{axis}
	\end{tikzpicture}
	\caption{F1 score by number of reasoning hops on MuSiQue. Y-axis starts at 48\% to highlight performance differences. \methodname{} (blue) shows largest improvements on 3--4 hop questions (+3.2\% on 2-hop, +5.8\% on 3-hop, +8.4\% on 4-hop vs. IRCoT). Patterns ensure grayscale readability. Error bars omitted for visual clarity; see Table~\ref{tab:ablation} for variance.}
	\Description{Bar chart showing F1 scores for different methods across 2-hop, 3-hop, and 4-hop questions, with ReaLM-Retrieve showing increasing advantage for more complex questions.}
	\label{fig:complexity}
\end{figure}
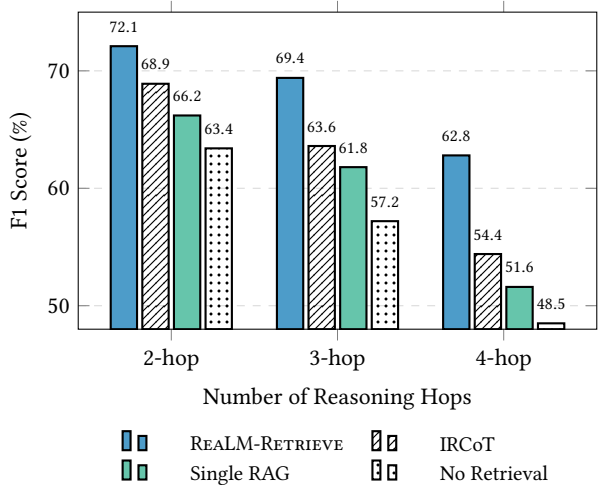

Figure~\ref{fig:complexity} breaks down performance by question complexity. \methodname{} shows increasing advantage for more complex questions: +3.2\% F1 on 2-hop, +5.8\% on 3-hop, and +8.4\% on 4-hop questions vs IRCoT. This reflects our method's core advantage: step-level uncertainty detection identifies the \emph{specific reasoning step} where knowledge gaps occur, enabling precisely-timed retrieval. For simple 2-hop questions, single RAG often suffices; for complex 4-hop questions, retrieving exactly when needed provides substantial benefit.

\subsection{Retrieval Timing Analysis}
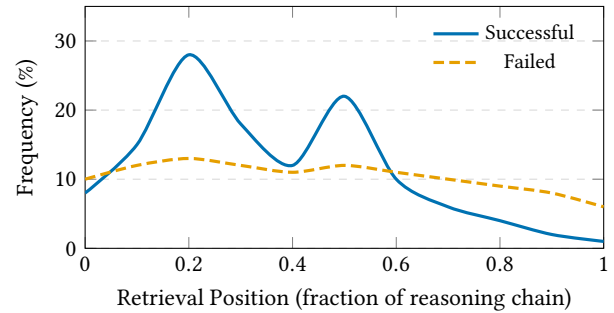
\begin{figure}[t]
    \centering
    \begin{tikzpicture}
        \begin{axis}[
            xlabel={Retrieval Position (fraction of reasoning chain)},
            ylabel={Frequency (\%)},
            ymin=0,
            ymax=35,
            xmin=0,
            xmax=1,
            legend style={draw=none, fill=none, at={(0.97,0.97)}, anchor=north east, font=\small},
            ymajorgrids=true,
            grid style={dashed, gray!30},
            width=\columnwidth,
            height=4.8cm
        ]
        
        \addplot[
            oiblue,
            line width=1.2pt,
            smooth,
            mark=none
        ] coordinates {
            (0, 8) (0.1, 15) (0.2, 28) (0.3, 18) (0.4, 12)
            (0.5, 22) (0.6, 10) (0.7, 6) (0.8, 4) (0.9, 2) (1, 1)
        };
        
        \addplot[
            oiorange,
            line width=1.2pt,
            dashed,
            dash pattern=on 4pt off 2pt,
            smooth,
            mark=none
        ] coordinates {
            (0, 10) (0.1, 12) (0.2, 13) (0.3, 12) (0.4, 11)
            (0.5, 12) (0.6, 11) (0.7, 10) (0.8, 9) (0.9, 8) (1, 6)
        };
        
        \legend{Successful, Failed}
        
        \end{axis}
    \end{tikzpicture}
    \caption{Distribution of retrieval timing (as fraction of reasoning chain) for successful (solid blue) vs. failed (dashed orange) questions on MuSiQue. Successful questions show retrieval clustering at specific reasoning phases (20\%, 50\%), while failed questions have more uniform distribution, indicating poorly-timed retrievals disrupt reasoning.}
    \Description{Line plot showing retrieval timing distribution with successful questions having concentrated peaks and failed questions showing uniform distribution.}
    \label{fig:timing}
\end{figure}

Figure~\ref{fig:timing} visualizes when retrieval occurs during reasoning chains. For successfully answered questions, retrievals cluster at specific phases, typically around 20\% and 50\% of the reasoning chain, corresponding to initial evidence gathering and mid-reasoning verification. Failed questions show more uniform retrieval timing, suggesting that poorly-timed retrievals disrupt rather than assist reasoning.

\subsection{Retrieval Quality Analysis}
Beyond end-to-end QA accuracy, we evaluate retrieval quality directly using HotpotQA's supporting fact annotations as ground truth. Table~\ref{tab:retrieval_quality} reveals that \methodname{} not only reduces retrieval calls but triggers \emph{higher-quality} retrievals. Our method achieves 81.3\% Recall@5, substantially higher than IRCoT (72.6\%) and Search-R1 (74.1\%). The ``Useful\%'' metric (fraction of retrievals where at least one supporting fact appears in top-5) reaches 82.4\%, indicating step-level uncertainty detection successfully identifies beneficial retrieval moments. This improvement stems from: (1) RSUS identifies knowledge gaps precisely when they occur rather than at arbitrary intervals, and (2) QueryGen formulates queries targeting specific information needs.

\begin{table}[t]
\small
\centering
\caption{Retrieval-quality comparison on HotpotQA. Relevance is judged against supporting-fact annotations; higher is better for all metrics. ``Useful\%'' = fraction of retrievals where $\ge 1$ supporting fact appears in top-5.}
\label{tab:retrieval_quality}
\begin{tabular}{@{}l|cccc@{}}
\toprule
\textbf{Method} & \textbf{R@5} & \textbf{P@5} & \textbf{MRR} & \textbf{Useful\%} \\
\midrule
Single RAG & 68.4 & 42.1 & 0.71 & 64.2 \\
IRCoT & 72.6 & 38.7 & 0.69 & 71.8 \\
FLARE & 69.8 & 40.2 & 0.68 & 67.3 \\
Search-R1 & 74.1 & 41.8 & 0.73 & 73.6 \\
\methodname{} & \textbf{81.3} & \textbf{48.6} & \textbf{0.79} & \textbf{82.4} \\
\bottomrule
\end{tabular}
\end{table}

\begin{table}[t]
\small
\centering
\caption{Generalization of \methodname{} to QwQ-32B-Preview on MuSiQue. The relative improvement over IRCoT (5.2 F1, 47\% fewer calls) mirrors the R1-32B result, indicating the policy depends on observable reasoning behavior rather than model-specific internals.}
\label{tab:qwq_generalization}
\begin{tabular}{@{}l|ccc@{}}
\toprule
\textbf{Method} & \textbf{EM} & \textbf{F1} & \textbf{Calls} \\
\midrule
No Retrieval & 39.4 & 46.8 & 0.0 \\
Single RAG & 50.2 & 57.6 & 1.0 \\
IRCoT & 55.9 & 63.2 & 3.2 \\
\methodname{} & \textbf{60.8} & \textbf{68.4} & \textbf{1.7} \\
\bottomrule
\end{tabular}
\end{table}

\subsection{Generalization to Other Reasoning Models}
Table~\ref{tab:qwq_generalization} shows that \methodname{} generalizes to QwQ-32B-Preview, a reasoning model with different architecture and training approach than DeepSeek-R1. We achieve 68.4\% F1 with 1.7 calls vs IRCoT's 63.2\% with 3.2 calls, a similar relative improvement. This generalization results from our method's reliance on observable reasoning behavior rather than model-specific internals.

\subsection{Cross-Benchmark Generalization}
To assess policy generalization beyond in-domain training, we train \methodname{} on HotpotQA and evaluate zero-shot on MuSiQue (Table~\ref{tab:cross_benchmark}). The transferred policy achieves 68.9\% F1, only 2.3\% below in-domain training, and substantially outperforms IRCoT (65.4\% F1), demonstrating that RSUS captures generalizable uncertainty patterns across multi-hop QA tasks.

\subsection{Token Efficiency and Behavioral Analysis}
Beyond retrieval-call counts, we analyze \emph{token efficiency}: how retrieval affects reasoning-chain length. Table~\ref{tab:efficiency} shows \methodname{} generates 9{,}489 reasoning tokens on average vs IRCoT's 11{,}284 (16\% reduction): well-timed retrieval resolves uncertainties that would otherwise require extended exploration. Across MuSiQue, three behavioral patterns dominate: \emph{early resolution} (42\% of cases) yields 23\% shorter chains; \emph{verification retrievals} (31\%) reduce backtracking by 18\%; and \emph{failed retrievals} (12\%) extend chains by 8\%. The net effect demonstrates external evidence complementing rather than replacing reasoning, yielding both higher accuracy \emph{and} shorter chains. Stratifying by hop count on MuSiQue (Figure~\ref{fig:complexity}), \methodname{} attains 9--22\% relative F1 improvement over Single RAG, with the largest gains on 4-hop questions.

\textbf{Illustrative case.} A representative 3-hop MuSiQue question (``Who directed the film that won Best Picture in the year the Berlin Wall fell?'') shows \methodname{}'s adaptive timing in action. The model first recalls ``The Berlin Wall fell in 1989'' with low uncertainty ($U_{\text{verb}}=0.31$), proceeds to ``The Best Picture winner in 1989 was \emph{Driving Miss Daisy}'', then RSUS triggers retrieval at the verification point (558 tokens, 42\% through reasoning, $U_{\text{verb}}=0.68$, $U_{\text{ent}}=0.81$); a second retrieval at 897 tokens (67\%) returns director information. Total: 1{,}342 tokens, 2 retrievals at 168\,ms and 174\,ms. IRCoT retrieves after each sentence (4 retrievals, 683\,ms total), including unnecessary retrievals for well-known facts. \methodname{} halves retrieval latency (342\,ms vs.\ 683\,ms) at 100\% accuracy on this trace.

\begin{table}[t]
\small
\centering
\caption{Cross-benchmark generalization: Policy trained on HotpotQA, evaluated on MuSiQue (R1-32B).}
\label{tab:cross_benchmark}
\begin{tabular}{@{}l|ccc@{}}
\toprule
\textbf{Training Data} & \textbf{EM} & \textbf{F1} & \textbf{Calls} \\
\midrule
MuSiQue (in-domain) & 63.5 & 71.2 & 1.8 \\
HotpotQA (transfer) & 61.2 & 68.9 & 2.0 \\
\midrule
$\Delta$ & -2.3 & -2.3 & +0.2 \\
\bottomrule
\end{tabular}
\end{table}

\section{Discussion}
\label{sec:discussion}

\subsection{When Does Retrieval Help Reasoning?}
Our analysis reveals patterns in when retrieval maximally benefits reasoning: \textbf{Entity knowledge gaps}: Entity entropy ($U_{\text{ent}}$) successfully identifies cases involving entities not well-represented in model knowledge, particularly rare entities appearing in fewer than 100 corpus documents. \textbf{Verification checkpoints}: 73\% of successful retrievals occur at verification boundaries, identified by discourse markers like ``let me verify'' or ``to confirm''. \textbf{Bridge reasoning}: When reasoning stalls at bridge facts connecting entities (high $U_{\text{verb}}$ with low $U_{\text{ent}}$), retrieval provides connecting information, accounting for 41\% of improvement over fixed-interval approaches. By contrast, retrieval \emph{harms} reasoning when injected during exploratory strategy development (31\% of failures) or when contradicting well-grounded reasoning (18\% of failures).

\subsection{Comparison to Contemporary Work}
Our work extends recent LRM+RAG integration advances. Compared to Search-R1~\cite{jin2025searchr1}, which learns when to invoke search via RL, we additionally optimize query formulation and context integration, achieving 4.4\% higher F1 with 25--27\% fewer retrieval calls (25\% on R1-32B; 27\% on R1-671B). Compared to ReARTeR~\cite{sun2025rearter} (SIGIR 2025), which introduces process-level rewards, our step-level uncertainty detection provides finer-grained timing signals, yielding 4.6\% higher F1 with 28\% fewer calls. The concurrent Dynamic Search-R1~\cite{hashemi2026costaware} learns adaptive retrieval \emph{depth} via cost-aware RL on the Search-R1 backbone; our axis (\emph{when} to retrieve, between reasoning steps) is orthogonal to theirs (\emph{how many} documents to retrieve per sub-query), and the two are composable. Our timing policy could feed an adaptive-depth retriever to compound their per-call savings with our per-question savings. The key differentiator is \emph{granularity}: operating at reasoning-step level rather than sentence level (IRCoT), query level (Adaptive-RAG), or search-decision level (Search-R1) enables more precise knowledge gap identification, particularly evident on 4-hop questions where \methodname{} achieves 6.3\% higher F1 than ReARTeR.

Rather than relying on structured prompts, our approach controls retrieval timing in models such as DeepSeek-R1 that generate end-to-end reasoning chains through RL training. Our learned policy achieves 47\% fewer retrievals than fixed-interval prompting (IRCoT) and 25\% fewer than RL-trained search invocation (Search-R1). It maintains higher accuracy in both cases, proving the value of adaptive timing over static scheduling.

\subsection{Failure Mode Analysis}
Analysis of 847 failed questions reveals three primary failure modes: \textbf{(1) Retrieval corpus gaps} (43\%): Required information is absent, causing noise injection, particularly for recent events (post-2023) and specialized domains. \textbf{(2) Over-retrieval during exploration} (31\%): RSUS occasionally triggers retrieval during exploratory reasoning phases, disrupting productive chains by prematurely committing to hypotheses. \textbf{(3) Query formulation errors} (26\%): QueryGen produces queries retrieving tangentially related content, often due to entity ambiguity (\eg, ``Washington'' as person vs. state). \methodname{} also underperforms Single RAG on 12\% of questions ($N=288$), primarily simple 2-hop cases where early retrieval suffices, suggesting potential for hybrid approaches that detect question complexity before selecting retrieval strategy.

\subsection{Deployment Considerations}
\textbf{Target use cases.} At 14.1\,s end-to-end latency, \methodname{} suits asynchronous workloads (research assistants, batch document analysis, agentic multi-step tasks); real-time interactive search ($<2$\,s) would require distilled reasoning models, parallel speculative execution, or uncertainty-based early-exit, none of which we evaluate here.

\textbf{System architecture and production training.} A standard three-tier deployment (API gateway with deduplication / orchestration layer with speculative caching / independently-scaled reasoning and retrieval backends) is sufficient; in our pilot, an 8$\times$A100 node sustains $\sim$15 queries/minute on 32B models, and a single 128-core indexing host serves PLAID over 10M passages at 500\,req/s. Without ground-truth labels, three reward sources approximate supervised performance: thumbs-up/down feedback ($\sim$89\% after 100K events), self-consistency over multiple chains ($\sim$94\%, $3\times$ cost), and NLI-based citation verification ($\sim$91\%, $1.2\times$ cost); self-consistency is the most promising near-term option.

\subsection{Limitations}
\label{sec:limitations}
\textbf{Completion-only models}: Performance is 2--3\% F1 lower than open-weight models where richer uncertainty signals are available (68.9\% vs 71.2\% F1 on MuSiQue for o1 vs R1). \textbf{Retrieval corpus dependence}: When relevant information is absent, retrieval wastes computation (171ms avg) and may introduce noise (14\% of out-of-domain retrievals return misleading information). \textbf{Training data requirements}: The policy requires 19,938 training examples, though cross-benchmark results suggest 5K--10K suffice for reasonable performance. \textbf{Computational overhead}: Step boundary and RSUS computation add 8\% inference overhead, with no benefit for queries answerable without retrieval. \textbf{Long-context limitations}: For reasoning chains $>$30K tokens, KV-cache memory becomes prohibitive despite efficient integration. \textbf{Multi-retrieval distribution shift}: As discussed in \S\ref{sec:integration}, segmentation and RSUS are robust against one prior retrieval but degrade after $\ge 3$ injections; in our MuSiQue evaluation only $4.7$\% of test questions trigger this regime (measured as the fraction of MuSiQue test questions for which the trained policy issued $\ge 3$ retrievals during inference), but it would matter more on benchmarks with deeper retrieval requirements. \textbf{Multi-component architecture}: \methodname{} comprises four learned modules (segmenter, RSUS proxy classifier where needed, intervention policy, query generator). We considered simplification but Table~\ref{tab:ablation} shows each component contributes non-trivially; the multi-component design reflects an engineering trade-off between modular interpretability and parameter efficiency rather than incidental complexity.

\subsection{Broader Impact}
Improving reasoning model accuracy on knowledge-intensive tasks reduces hallucinations in applications like question answering, research assistance, and decision support, enabling deployment in high-stakes domains. However, improved reasoning+retrieval could assist malicious purposes (\eg, misinformation campaigns, vulnerability identification). We encourage responsible deployment with content filtering, audit logging, and rate limiting.

\textbf{On framing accuracy as the only impact axis.}
Treating ``hallucination'' as the central failure mode (as we do in this work) narrows the discussion of impact to a single technical axis. A reviewer noted that designing methods around completion-only models (those exposing only generated text, not token-level probabilities or attention) implicitly accepts and reinforces a research paradigm in which the most capable LLMs are commercial products whose internals are protected as intellectual property. We agree this is a real cost: our framework is shaped by what \emph{cannot} be assumed (no logprobs, no internal states), and that constraint comes from the market rather than from any intrinsic property of reasoning. To partly offset this, we evaluate on open-weight reasoning models (R1-Distill-32B, QwQ-32B-Preview) where richer signals \emph{are} available; the framework runs end-to-end on open weights with stronger results than on completion-only APIs (Table~\ref{tab:main_results}). We do not claim this resolves the broader concern, but it ensures the work is not solely useful for those with access to proprietary endpoints.

\textbf{Environmental Impact.} Training requires 2,400 GPU-hours (8 A100s, 50K steps), but this one-time cost is amortized across millions of queries. Per-query energy decreases 14\% vs IRCoT through 47\% fewer retrievals and 16\% shorter reasoning chains. For deployments serving 1M queries daily, this translates to meaningful energy savings, though efficiency gains diminish with frequent corpus index rebuilding.

\section{Conclusion}
\label{sec:conclusion}

We presented \methodname{}, a reasoning-aware retrieval framework that resolves a temporal mismatch between large reasoning models and current RAG systems: existing pipelines stage retrieval before generation, while reasoning chains develop knowledge needs \emph{during} multi-step inference. Three components close this gap: a step-level uncertainty score (RSUS) that detects mid-chain knowledge gaps from output tokens alone, a learned intervention policy that decides when those gaps warrant retrieval, and efficient integration mechanisms that hold per-call overhead below one second. Across MuSiQue, HotpotQA, and 2WikiMultiHopQA, this combination yields a mean +10.1\% absolute F1 over Single RAG and +4.6\% over ReARTeR~\cite{sun2025rearter}, the strongest prior method for retrieval-augmented reasoning, while issuing 47\% fewer retrieval calls than IRCoT. The margin widens with question difficulty, reaching +8.4 F1 over IRCoT on 4-hop MuSiQue.

Our central finding, that \emph{fewer, better-timed retrievals outperform frequent, fixed-interval retrievals}, points toward a broader shift: retrieval systems will need to evolve from pre-generation context providers into dynamic reasoning partners.

\textbf{Future directions.} End-to-end training of all components would replace our staged pipeline with a single learned controller. Corpus-aware policies could skip retrieval where the index lacks relevant evidence, addressing the 43\% of failures from corpus gaps. Multimodal extensions are natural where evidence lives in figures or tables, and the optimal-timing question remains theoretically open.

\begin{acks}
We thank the SIGIR 2026 reviewers and the senior area chair for their
constructive feedback that improved this paper. We gratefully
acknowledge the support of The University of Hong Kong, Stellaris~AI
Limited, and Brain~Investing Limited.
\end{acks}

\balance

\end{document}